\documentclass[12pt]{iopart}

\usepackage{graphicx,color}
\usepackage{multirow}

\begin{document}

\title{Magnetic field and contact resistance dependence of non-local charge imbalance}

\author{A Kleine$^1$, A Baumgartner$^1$, J Trbovic$^1$, D S Golubev$^2$, A D Zaikin$^{2,3}$ and C Sch\"onenberger$^1$}

\address{$^1$
Institute of Physics, University of Basel, Klingelbergstrasse 82, 4056 Basel, Switzerland\\}

\address{$^2$ Institute for Nanotechnology, Karlsruhe Institute of Technology (KIT), 76021 Karlsruhe, Germany\\}

\address{$^3$ I E
Tamm Department of Theoretical Physics, P N Lebedev Physics
Institute, 119991 Moscow, Russia\\}

\ead{andreas.baumgartner@unibas.ch}

\date{\today}

\begin{abstract}
Crossed Andreev reflection (CAR) in metallic nanostructures, a possible basis for solid-state electron entangler devices, is usually investigated by detecting non-local voltages in multi-terminal superconductor/normal metal devices. This task is difficult because other subgap processes may mask the effects of CAR. One of these processes is the generation of charge imbalance (CI) and the diffusion of non-equilibrium quasi-particles in the superconductor. Here we demonstrate a characteristic dependence of non-local CI on a magnetic field applied parallel to the superconducting wire, which can be understood by a generalization of the standard description of CI to non-local experiments. These results can be used to distinguish CAR and CI and to extract CI relaxation times in superconducting nanostructures. In addition, we investigate the dependence of non-local CI on the resistance of the injector and detector contacts and demonstrate a quantitative agreement with a recent theory using only material and junction characteristics extracted from separate direct measurements.

\end{abstract}

\pacs{73.23.-b, 74.45.+c, 74.78.Na, 03.67. Mn}

\maketitle


\section{Introduction}

Transport phenomena in superconductors have been heavily investigated for many years. Recently, sub-micron scaled multi-terminal normal metal/superconductor (NS) devices came to renewed attention of both, experimental \cite{Byers_PRL74_1995, Beckmann_PRL93_2004, Russo_Klapwijk_PRL95_2005, Cadden-Zimansky_Chandrasekhar_PRL97_2006, Cadden-Zimansky_NewJPhys9_2007, Cadden-Zimansky_Chandrasekhar_Nature_2009, Kleine_Baumgartner_EPL87_2009} and theoretical \cite{Recher_Loss_PRB63_2001,Lesovik_EPJB24_2001, Falci_EPL54_2001, Recher_PRL91_2003, Kalenkov_PRB75_2007, Yeyati_naturephysics3_2007, Golubev_Zaikin_PRL103_2009} investigations, mainly because of a quantum mechanically non-local process called crossed Andreev reflection (CAR). In {\it local} Andreev reflection, an electron impinging on the NS interface from the N side can enter the superconductor at subgap energies only by forming a Cooper pair (CP) with a second electron. To maintain momentum conservation a hole is retro-reflected into the N contact. As depicted in figure~1a, in CAR the hole is reflected into a second contact N2 separated by less than the superconducting coherence length from the first contact N1. The inverse process can be described as splitting of a Cooper pair with the electrons entering two separate contacts. Since ideally the two electrons retain their correlations from the superconductor, this process might provide a source of spin-entangled electron pairs in a solid-state environment \cite{Recher_Loss_PRB63_2001, Lesovik_EPJB24_2001, Recher_PRL91_2003}.

Cooper pair splitting has been demonstrated very recently in a double quantum dot system \cite{Hofstetter_Czonka_Schoenenberger_Nature_2009}, where the strong Coulomb interaction in the quantum dots leads to a suppression of other competing transport mechanisms. In contrast, the experimental observation of CAR in metallic structures is more challenging because of other subgap processes of similar probability, namely elastic co-tunnelling (EC) and non-local charge imbalance (CI). EC is the direct electron transfer from one contact to the other via virtual intermediate states in the superconductor. It has been shown theoretically that without interactions and to the lowest order in the tunnelling rate, EC exactly cancels the effects of CAR. Several recent publications \cite{Beckmann_PRL93_2004, Russo_Klapwijk_PRL95_2005, Cadden-Zimansky_Chandrasekhar_PRL97_2006,  Cadden-Zimansky_Chandrasekhar_Nature_2009, Kleine_Baumgartner_EPL87_2009} show experimentally that this cancellation can be lifted. Various models have been proposed to understand these findings \cite{Kalenkov_PRB75_2007, Yeyati_naturephysics3_2007, Golubev_Zaikin_PRL103_2009}, for example dynamical Coulomb blockade (DCB) and interactions via electromagnetic excitations of the environment \cite{Yeyati_naturephysics3_2007}. Here we present data that support our previous interpretation in terms of DCB \cite{Kleine_Baumgartner_EPL87_2009}. In the main part of the paper, however, we focus on charge imbalance (CI), i.e. on non-equilibrium effects in the superconductor. It is well-established that at finite bias charged quasi-particles (QPs) are excited above the gap in the superconductor by tunnelling injection of electrons, which leads to CI in the superconductor \cite{Tinkham_PRB6_1972}, see figure~1b. These QPs diffuse away from the injector contact and lead to a potential gradient in the superconductor, see figure~1c. Distinguishing CAR from CI and EC is essential for the investigation of CAR and can be achieved by ferromagnetic contacts \cite{Beckmann_PRL93_2004}, the sign of the non-local signal \cite{Russo_Klapwijk_PRL95_2005, Kleine_Baumgartner_EPL87_2009}, or the characteristic temperature \cite{Kleine_Baumgartner_EPL87_2009} or distance dependence \cite{Beckmann_PRL93_2004, Russo_Klapwijk_PRL95_2005, Cadden-Zimansky_Chandrasekhar_PRL97_2006}. Usually, several of these criteria have to be employed to discern the subgap processes. We show that non-local CI exhibits a strong characteristic dependence on a magnetic field applied parallel to the axis of the superconducting wire. This dependence is due to the reduction of the pair breaking time, well-described by the conventional CI description \cite{Tinkham_PRB6_1972, Schmid_Schoen_JLowTPhys20_1975} adapted to our non-local measurements. In addition, we study the contact resistance dependence of non-local CI and compare the results to a recent theoretical model of subgap processes in superconductors \cite{Golubev_Zaikin_PRL103_2009}.

\section{Sample characterization and non-local resistance}

\begin{figure}[b]{
\centering
\includegraphics{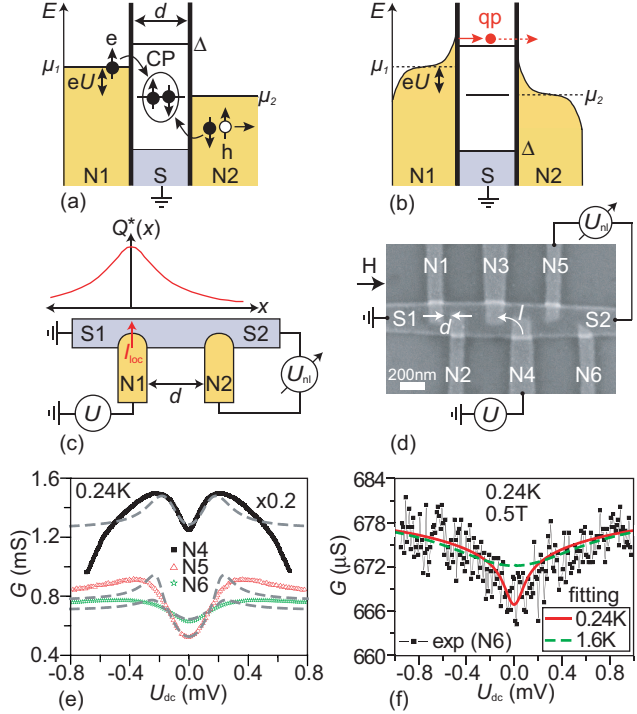}
}
\caption{(Color online) Schematic of the non-local processes
a) crossed Andreev reflection and b) charge imbalance. c)
schematic of the spatial distribution of quasi-particle density $Q^*(x)$ in a
non-equilibrium superconductor device. d) SEM image of the
sample. e) Barrier characteristics of barriers N4, N5 and N6. The gray
dashed lines are BTK fits with an additional broadening parameter. f) Normal state junction conductance of N6 at base
temperature at a magnetic field of $0.5\,$T$>H_{\rm c}$, including the finite
resistance of the Al leads. The lines without symbols show
the theoretical zero-bias anomaly due to dynamical Coulomb
blockade at the indicated temperatures.}
\end{figure}

The sample was fabricated on a thermally oxidized silicon substrate by electron beam lithography and UHV angle evaporation. It consists of a $250\,$nm wide aluminium wire with several palladium contacts fabricated on top of the tunnel barrier prepared by oxidation of the aluminum for 15~minutes in a $0.1\,$mbar oxygen atmosphere. The structure is shown in figure~1d. The aluminum is about $45\,$nm thick and has a resistivity of $7.0\,\mu\Omega$cm, a mean-free path of $8.5\,$nm and a `dirty limit' coherence length of $110\,$nm. In direct transport experiments we find a critical temperature of $1.3\,$K and a critical magnetic field of $240\,$mT (at $T=0.24\,$K and with the field applied parallel to the superconducting wire). We concentrate on data obtained using the three contacts N4, N5 and N6, with the conductance vs. bias characteristics shown in figure~1e. While the junctions N5 and N6 have similar conductances, N4 exhibits roughly five times larger values. The characteristics can be fitted to the BTK model \cite{Blonder_PRB25_1982}, with an additional broadening parameter and the barrier strength $Z$ used as fit parameters (for N4 we also slightly adjusted the $1.5\,$K normal state resistance $R$). The superconducting energy gap $\Delta=190\,\mu$eV and the temperature $T=0.24\,$K are not used for fitting.  $Z$ and $R$ are given in Table 1, together with the resistance-area product $RA$ of each junction. All barriers are in the regime of moderate transmission, ideal for the observation of CAR in our structures \cite{Kleine_Baumgartner_EPL87_2009}. Figure~1f shows the normal state resistance of contact N6 at base temperature and a magnetic field $H=0.5\,$T$>H_{\rm c}$ applied parallel to the superconductor wire and will be discussed below. The edge-to-edge distances between the contacts are $d_{45}=105\,$nm, $d_{56}=140\,$nm and $d_{46}=375\,$nm. The measurements were performed in a $^3$He cryostat with a base temperature of $0.24\,$K.

\begin{table}[t]
\centering
\caption{Contact properties. $R$: barrier resistance at $T=1.5\,$K; $A$: contact area; $RA$: resistance-area product; $Z$: BTK barrier strength; $g_{\rm NS}=G_{\rm NS}/G_{\rm NN}$: normalized tunnelling conductance at zero bias and $T=1.1\,$K.}

\begin{tabular}{cccccc}\\\hline\hline
 			
    & $R$  	& $A$ & $RA$	& $Z$ &  $g_{\rm NS}$ \\
	  &[k$\Omega$]& $[\mu{\rm m}^2$] & [$\Omega\mu{\rm m}^2$] & & (@$1.1\,$K) \\\hline												 
N4  & 0.20  & 0.032 & 6.4 	&  0.55 &  1.30 \\

N5  & 1.28  & 0.021 & 26.9  & 0.77  &  0.95\\

N6  & 1.40  & 0.007 & 10.4  & 0.61  &  1.01

 \\\hline\hline

	\end{tabular}
\end{table}

We measure the non-local resistance as shown schematically in figures~1c and 1d:  an ac modulated voltage $U_{\rm dc}+U_{\rm ac}$ with $U_{\rm ac}=12\,\mu$Vrms is applied to a normal metal contact (injector), e.g. N4 in figure~1d, which leads to a local current $I_{\rm dc}+I_{\rm ac}$ through the superconducting wire to ground via S1. We record the ac-modulation $U_{\rm nl}$ of the non-local voltage between the normal contact N5 (detector, not in the current path) and the superconductor at S2, more than $10\mu$m away from the normal contacts, by lock-in techniques at a modulation frequency of $\sim10\,$Hz. The non-local differential resistance can be approximated by $R_{\rm nl}=U_{\rm nl}/I_{\rm ac}$. We expect $R_{\rm nl}<0$ for CAR and $R_{\rm nl}>0$ for EC and CI \cite{Kleine_Baumgartner_EPL87_2009}.

On the present device we obtain similar $R_{\rm nl}$~vs.~$U_{\rm dc}$ traces as reported previously \cite{Kleine_Baumgartner_EPL87_2009}. As an example, a series of curves at different temperatures is shown for the injector-detector pair (N6, N5) in figure~2a. At base temperature $R_{\rm nl}$ exhibits a local maximum at zero bias and becomes negative at finite subgap voltages. $R_{\rm nl}$ is positive for bias potentials $eU>\Delta$. These characteristics remain unchanged up to $T=0.5\,$K. At higher temperatures the subgap feature is washed out and the signal becomes positive for all voltages. Similar curves are obtained for the contact pair (N4, N5), except that no zero-bias maximum develops (not shown). We attribute a negative non-local resistance to CAR and a positive sign to EC or CI. The development of the local maximum at low temperatures for the data with the less transparent injector N6 is probably due to a shift in the relative strength of CAR and EC, caused by enhanced dynamical Coulomb blockade \cite{Kleine_Baumgartner_EPL87_2009}. 

To further support this interpretation we report here a zero-bias anomaly in the conductance of the junctions N5 and N6, consistent with dynamical Coulomb blockade. As an example, we plot the conductance of junction N6 at base temperature in figure~1f. Here, superconductivity is suppressed by an external magnetic field of $0.5\,$T, to avoid features due to the superconductor gap. We find a weak, but clear reduction of the junction conductance at zero bias. We also plot the theoretically expected response due to dynamical Coulomb blockade for a junction capacitance of $0.5\,$fF (estimated from a plate capacitor model) with a resistive environment \cite{Ingold_Nazarov}. The only fit parameter is the environmental resistance, which we adjusted to $105\,\Omega$. This value might be explained by the transmission-line response of the cryostat cables. The zero-bias anomaly is strongly suppressed at higher temperatures. We observe a similar feature in the conductance of N5, but a constant conductance for N4. When the latter is used as injector we observe no maximum at zero bias in the non-local signal. 

\section{Magnetic field dependence of CI}

Now we discuss the magnetic field dependence of the subgap signals. From direct conductance measurements of the aluminum strip we infer a critical field of $H_{\rm c}\approx 240\,$mT. In figure~2b, a series of $R_{\rm nl}$ vs. $U_{\rm dc}$ curves of the contact pair (N6, N5) is shown for different magnetic fields applied parallel to the Al wire. At bias potentials larger than the superconductor energy gap the non-local resistance is strongly reduced already for relatively small fields, $H<<H_{\rm c}$. In contrast to this, the subgap features are not or only very weakly affected. At $H>H_{\rm c}$, e.g. for $H=500\,$mT, the non-local response becomes zero for all voltages.

These findings can be seen clearer in magnetic field traces for the contact pair (N4, N5), which exhibits an appreciable non-local resistance at base temperature and zero bias, see figure~2c. At $T=0.24\,$K the signal is negative and we tentatively attribute it to CAR. We find that this signal is independent of $H$ for $H<<H_{\rm c}$. Only near $H_{\rm c}$ the signal increases and reaches positive values. We note that the signal does not reach zero exactly for $H>H_{\rm c}$, possibly due to a small cross talk current in the device, though the signal does not change upon doubling of the measurement frequency. In contrast to measurements at base temperature, $R_{\rm nl}$ is positive for $T=1.1\,$K and $H=0$ and is reduced strongly for small fields. Near $H_{\rm c}$ the signal increases strongly. For $H>H_{\rm c}$ the signal is essentially zero. We obtain a similar magnetic field trace when we apply a finite bias to the injector, as shown in figure~2d for $U_{\rm dc}=0.2\,$mV and $U_{\rm dc}=0.3\,$mV at base temperature, with critical fields of $205\,$mT and $190\,$mT, respectively, and an enhanced signal strength due to the larger injected QP currents.

\begin{figure}[t]{
\centering
\includegraphics{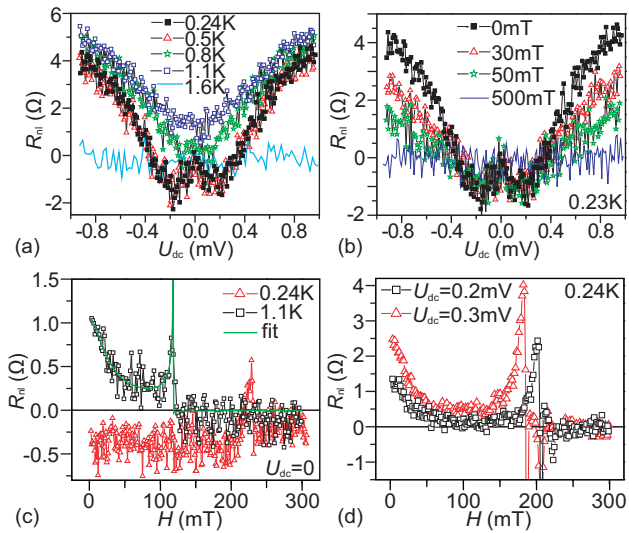}
}
\caption{(Color online) Non-local resistance vs. bias for a) a series of temperatures and b) a series of magnetic fields parallel to the Al strip for the contact pair (N6, N5). c) Magnetic field traces of $R_{\rm nl}$ at zero bias at $T=0.24\,$K and $T=1.1\,$K for the contact pair (N4, N5). The green line is a fit as discussed in the text. d) Magnetic field dependence of $R_{\rm nl}$ at finite bias and base temperature for the contact pair (N4, N5).}
\end{figure}

Our magnetic field data on CAR are in stark contrast to previous experiments \cite{Russo_Klapwijk_PRL95_2005}, where CAR was found to be strongly suppressed by a magnetic field parallel to the superconducting film. This inconsistent behaviour might be related to the different coherence lengths in the materials used for the superconductor wire and normal metal contacts, or the different sample geometries. However, we note that in contrast to our previous experiments \cite{Kleine_Baumgartner_EPL87_2009}, the out-of-phase part of the low-temperature signal in figure~2c has a similar amplitude as the in-phase signal, so that our results on CAR have to be considered as preliminary.

In contrast to CAR, the magnetic field dependence of CI can be well understood by adapting the standard CI description \cite{Tinkham_PRB6_1972} to non-local experiments: we assume that the injected quasi-particles (QPs) diffuse isotropically in the Al wire (ideally, the electric field in the superconductor is zero) and that the QP concentration near the detector contact determines the induced voltage in a similar way as in local measurements. This leads to the following expression for the non-local resistance:

\begin{equation}
R_{\rm nl}=\frac{F^*(U_{\rm dc}) \tau_{Q^*}}{2e^2N\Omega_{\rm inj} g_{\rm NS,det}(0)}e^{-\left|x\right|/\Lambda_{Q^*}},
\label{Rnl}
\end{equation}
where $\tau_{Q^*}$ denotes the CI relaxation time, $F^*(U_{\rm dc})$ the fraction of current that is carried by the injected QPs at a given bias, $N$ the density of states of the normal state Al, $\Omega_{\rm inj}$ the injection volume, for which we insert the superconductor volume below the injector contact. $g_{\rm NS,det}(0)$ is the zero bias detector conductance at the temperature of the experiment, normalized to the conductance of the junction in the normal state, and $x$ is the distance between injector and detector. The charge imbalance relaxation length $\Lambda_{Q^*}$ is the distance over which the QP distribution function in a superconductor relaxes to its equilibrium value. It is related to $\tau_{Q^*}$ by $\Lambda_{Q^*}+\sqrt{D\tau_{Q^*}}$, with $D$ the
electron diffusion constant in Al. Several processes influence $\tau_{Q^*}$, like inelastic electron-phonon scattering, pair breaking by spin-active impurities, orbital motion, a finite supercurrent or inhomogeneities in the order parameter $\Delta$. Generally, the QP life time reads \cite{Schmid_Schoen_JLowTPhys20_1975, Stuivinga_Klapwijk_JLowTPhys53_1983}

\begin{equation}
\tau_{Q^*}=\frac{4 k_{\rm B}T}{\pi\Delta(T,H)}\sqrt{\frac{\tau_{\rm E}}{2\Gamma}}.
\end{equation}

Around zero bias the injector current is small and we can ignore effects due to pair breaking of the supercurrent. For a homogeneous gap parameter this leads to  $\Gamma=\tau_{\rm S}^{-1}+(2\tau_{\rm E})^{-1}$ with $\tau_{\rm S}$ the orbital pair breaking and $\tau_{\rm E}$ the inelastic scattering time. For the temperature dependence of the energy gap we use the standard BCS expression $\Delta(T)=1.74\Delta(0)\sqrt{1-T/T_{\rm c}}$ and for the magnetic field dependence $\Delta(T,H) = \Delta(T)\sqrt{1-H^2/H_{\rm c}(T)^2}$ with $H_{\rm c}(T)=H_{\rm c}(0)(1-T^2/T_{\rm c}^2)$. The orbital pair breaking time as a function of $H$ is given by $\tau_{\rm S}=\hbar/\Delta(0, 0)\cdot H_{\rm c}^2/H^2$ \cite{Stuivinga_Klapwijk_JLowTPhys53_1983}.

At zero bias the only fit parameters are $F^*$ and $\tau_{\rm E}$ and we obtain $F^*=0.05$ and $\tau_{\rm E}=0.25\,$ns, corresponding to $\Lambda^*=1.0\,\mu$m, for the fit shown in figure~2c. As expected, $F^*<<1$ since the total current at zero bias is dominated by Andreev reflection. The relaxation time is considerably smaller than reported for thick ($\sim12\,$ns) and thin films ($\sim4\,$ns) \cite{Stuivinga_Klapwijk_JLowTPhys53_1983}, but consistent with the reduced charge imbalance length found in other experiments on superconducting Al wires \cite{Cadden-Zimansky_Chandrasekhar_PRL97_2006, Cadden-Zimansky_NewJPhys9_2007}. The reduction of $\tau_{\rm E}$ for thin films was attributed to enhanced electron-electron scattering for films of thickness smaller than $\sqrt{\hbar D/k_{\rm B}T}$ \cite{Stuivinga_Klapwijk_JLowTPhys53_1983}, which we estimate to $\sim 160\,$nm at $T=1.1\,$K for our structures. We expect a similar additional suppression in thin wires.

The finite bias experiments shown in figure~2d were performed at base temperature, which leads to a larger energy gap and a characteristic increase of CI at a larger field compared to figure~2c. In addition, the finite supercurrent at an increased bias leads to a reduction of the energy gap and we expect a reduction of the QP relaxation time due to additional pair breaking. The latter effect is strong only for large currents near $H_{\rm c}$, so that we can attribute the increase in $R_{\rm nl}$ with bias at $H=0$ to a change of $F^*$: for a larger bias a larger fraction of the total current is generated by QP injection. We note that $R_{\rm nl}$ turns negative when the injected current is comparable to the critical current for given external parameters. We observe this effect in figure~2d for $H\geq H_{\rm c}$, as well as in the bias sweeps at elevated temperatures shown below in figure~3a. We attribute this phenomenon to the initial stages of the breakdown of superconductivity, possibly at phase-slip centers, which renders the current inhomogeneous.

\section{Contact resistance dependence of charge imbalance}

The previous section shows that the QP diffusion is well described by the standard theory. 
We now turn our attention to the contact resistance dependence of the CI induced non-local resistance. 
This is an important question since CAR and EC are strongly affected by the junction characteristics 
\cite{Kleine_Baumgartner_EPL87_2009}.

\begin{figure}[b]{
\centering
\includegraphics{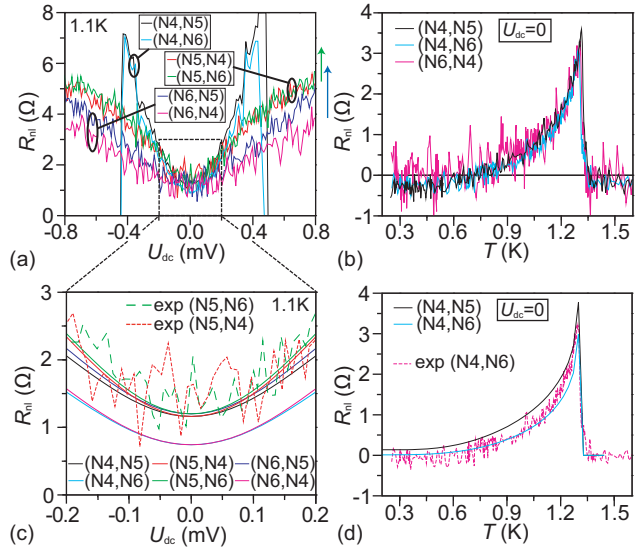}
}
\caption{(Color online) Non-local resistance a) vs. bias at $T=1.1\,$K and b) vs. temperature at zero 
bias for the indicated injector-detector pairs. c) and d) show the corresponding curves calculated in 
the model of reference \cite{Golubev_Zaikin_PRL103_2009} with the experimental traces discussed in the text.}
\end{figure}

To obtain a reasonable signal strength, we performed the experiments at $T=1.1\,$K. In figure~3a the non-local resistance $R_{\rm nl}$ is plotted vs. bias for a series of injector-detector pairs. 
The curves with the same injector contact are almost identical and can be clearly distinguished from the other curves. 
The distinction is most obvious if we use N4 as injector, with a considerably larger conductance than the other contacts. 
In contrast, using N4 as detector does {\it not} lead to a significant change in the non-local resistance, as can be seen 
for example for the pairs (N5, N4) and (N5, N6). A comparison of the non-local resistance at zero bias between injector 
and detector pairs as a function of temperature is shown in figure~3b. We note the characteristic shape of the curve that 
allows one to identify CI \cite{Cadden-Zimansky_Chandrasekhar_PRL97_2006, Cadden-Zimansky_NewJPhys9_2007}. We find that 
near zero bias and for $T>0.5\,$K the non-local resistance becomes the same for all contact pairs and thus seemingly 
independent of the contact resistances. 
For bias potentials larger than the energy gap the corresponding curves are not identical and the signal amplitude depends 
stronger on the injector than on the detector resistance, see figure 3a.

First we compare our data with the standard theory described by equation~1 and investigate the relative change 
between two curves. For the measurements with the same injector, $F^*$ and $\Omega_{\rm inj}$ are identical and 
we expect that only the distance between the contacts and the normalized conductance $g_{\rm NS,det}$ are relevant. 
As an example, using the experimentally determined $g_{\rm NS,det}$ at $T=1.1\,$K and the CI length obtained in the 
previous section, we expect a factor of $\sim1.3$ between data points from the pairs (N5, N4) and (N5, N6) at the same bias. In figure~3c these two curves are replotted on a smaller scale. We obtain a factor $1.0\pm0.3$ between these two curves at zero bias. The standard deviation is even larger for the other pairs. We therefore conclude that the deviation from the standard description of CI is not significant due to the low signal-to-noise ratio. At higher bias the deviation from equation~1 is more pronounced, as shown in figure~3a, where the expected non-local resistance at $U_{\rm dc}=0.8\,$mV is indicated by a green and a blue arrow for the pairs (N5, N6) and (N6, N5), based on the values of the pairs (N5, N4) and (N6, N4), respectively. However, in this regime an appreciable supercurrent is flowing in the device and we do not expect equation~1 to hold exactly.

Swapping the injector and detector contacts allows one to compare signals independent of the contact distance. In this scheme, however, it is necessary to estimate $F^*$ and the injection volume, each introducing considerable errors. 
For example, by assuming that $F^*$ is similar for two contacts (e.g. for large bias $F^*\approx1$) one finds a factor of $\sim3.5$ between the curves with swapped contacts (N4, N6) and (N6, N4). This is not supported by our data, see figure~3b, which shows that $F^*$ varies between the contacts even at zero bias.

Our experimental results can be understood quantitatively by the model put forward in
reference \cite{Golubev_Zaikin_PRL103_2009}, which we adapt slightly so that the experimental junction parameters in table~1 can be used. The non-local resistance of our device in terms of conductances reads
\begin{eqnarray}
R_{\rm nl}(U_{\rm dc})=\frac{G_{\rm nl}(U_{\rm dc})}{G_{\rm det}(0)G_{\rm inj}(U_{\rm dc})-G_{\rm nl}^2(U_{\rm dc})},
\label{Rnl1}
\end{eqnarray}
where
\begin{eqnarray}
G_{\alpha}(U_\alpha)=\frac{1}{R_\alpha}\int dE \frac{g_\alpha(E)}{4k_{\rm B}T\cosh^2\frac{E-eU_\alpha}{2k_{\rm B}T}}
\end{eqnarray}
($\alpha=$inj, det) is the local conductance of the injector and detector and
$g_\alpha(E)$ are the energy dependent spectral conductances of the NS barriers
\cite{Blonder_PRB25_1982}: 
\begin{eqnarray}
g_\alpha(E)&=&\frac{2\theta(\Delta-|E|)(1+Z_\alpha^2)\Delta^2}{E^2+(\Delta^2-E^2)(1+2Z_\alpha^2)^2}
\nonumber\\ &&
+\, \frac{2\theta(|E|-\Delta)(1+Z_\alpha^2)|E|}{|E|+\sqrt{E^2-\Delta^2}(1+2Z_\alpha^2)}.
\end{eqnarray}
with $\theta$ the Heaviside function. The general expression for the non-local conductance $G_{\rm nl}(U_{\rm dc})$
is presented in reference \cite{Golubev_Zaikin_PRL103_2009}. Provided the
charge imbalance length $\Lambda_{Q^*}$ is shorter than the length of a
superconducting wire (which is the case in our experiments), we find
\begin{eqnarray}
&& G_{\rm nl}(U_{\rm dc})=\frac{1}{4e^2NDS R_{\rm inj}R_{\rm det}}
\nonumber\\ &&\times\,
\int\limits_{|E|<\Delta} dE \frac{g_{\rm inj}(E)g_{\rm det}(E)}{4k_{\rm B}T\cosh^2\frac{E-eU_{\rm dc}}{2k_{\rm B}T}}
\frac{\Delta^2-E^2}{\Delta^2}\frac{e^{-k(E)|x|}}{k(E)}
\nonumber\\ &&
+\,\frac{\Lambda_{Q^*} e^{-|x|/\Lambda_{Q^*}}}{4e^2NDSR_{\rm inj}R_{\rm det}}
\nonumber\\ &&\times\,
\int\limits_{|E|>\Delta} dE \frac{g_{\rm inj}(E)g_{\rm det}(E)}{4k_{\rm B}T\cosh^2\frac{E-eU_{\rm dc}}{2k_{\rm B}T}}
\frac{E^2-\Delta^2}{E^2},
\end{eqnarray}
where $k(E)=\sqrt{{2\sqrt{\Delta^2-E^2}}/{D}+{1}/{\Lambda_{Q^*}^2}}$
and $S$ is the superconducting wire cross-section. 
Note that at zero bias and in the limit $\Delta\ll k_{\rm B}T$, i.e. close to the critical temperature
or the critical magnetic field, 
equation \ref{Rnl1} reduces to equation \ref{Rnl} with $F^*=\Omega_{\rm inj}/2S\Lambda_{Q^*}$.
One can see that for $G_{\alpha}\gg G_{\rm nl}$ the resistances of the NS barriers cancel out and do not enter the expression for the non-local resistance, which depends on the barrier properties
only through the BTK parameters $Z_\alpha$. Furthermore, in both limits of high and
low temperatures also these factors
drop out and one arrives at the following simple expressions for the non-local resistance    
\begin{eqnarray}
R_{\rm nl}(0)=\left\{ 
\begin{array}{ll}
(r_{\Lambda_{Q^*}}/2)\,e^{-|x|/\Lambda_{Q^*}}, & T_C-T\ll T_C, \\
(r_{\xi}/2)\,e^{-|x|/\xi}, & k_{\rm B}T\ll \Delta,
\end{array}
\right.
\label{Rnlf}
\end{eqnarray}
where $r_{\Lambda_{Q^*}}$ and $r_\xi$ are the normal state resistances of the
superconducting wire segments of lengths $\Lambda_{Q^*}$ and $\xi$, respectively.
Thus, we find that the non-local resistance in these limits only depends
on the distance between the junctions and the
properties of the superconducting wire, 
but not on the properties of the NS barriers.

In figure~3c, numerically calculated $R_{\rm nl}$ 
curves based on the above equations are shown for $T=1.1\,$K, with material and contact characteristics from the experiments, including $\Lambda_{\rm Q*}=1\,\mu$m from the magnetic field dependence. We stress that no fit parameters are used for these plots. We find that the calculations quantitatively reproduce our data. The two curves with larger contact separation exhibit a reduced non-local resistance compared to the other curves, which could not be resolved in the experiment. Figure~3d shows calculated temperature sweeps, which compare very well to the experimental curves. For this comparison we have subtracted a small constant from the data, since the model does not account for negative non-local resistance, because no electron interactions are considered in this model. In this model, the swapping of the injector and detector results in identical curves, also in agreement with our data.

\section{Conclusions}
In summary, we report measurements of the non-local resistance in a multi-terminal superconductor/normal metal device. We show a zero-bias anomaly at base temperature in the normal state of the superconductor consistent with dynamical Coulomb blockade, which might be partly responsible for lifting the balance between crossed Andreev reflection (CAR) and elastic co-tunnelling. We then focus on charge imbalance (CI) and demonstrate a characteristic magnetic field dependence, which provides a novel tool to distinguish CAR and CI in future experiments.
By comparing our data to a generalization of the standard description of CI, we can extract the inelastic quasi particle relaxation time and find a reduction relative to literature values probably due to the reduced 
dimensions of the superconductor wire.

In addition we present a set of experiments where we analyze systematically pairs of injector and detector 
contacts. To generate CI, we chose a rather high temperature, at which the superconducting and the normal junction resistances differ only weakly, which results in only small differences in the curves. The results suggest that the contact resistances influence the non-local CI only weakly. This finding is reproduced in a theoretical model adapted to incorporate experimentally accessible parameters. This model does not require additional adjustable parameters and agrees quantitatively with our data.

\ack
This work is financially supported by the Swiss National Science Foundation
and in part by DFG and by RFBR under grant 09-02-00886.

\end{document}